\documentclass[12pt]{article}
\usepackage{amssymb,amsmath,bm,epsfig,comment}

\renewcommand{\thefootnote}{\fnsymbol{footnote}}
\newcommand{\im}{{\rm Im\,}}
\newcommand{\re}{{\rm Re\,}}

\begin{document}
\title{}

\title{
\begin{flushright}
\begin{minipage}{0.2\linewidth}
\normalsize
WU-HEP-15-18 \\*[50pt]
\end{minipage}
\end{flushright}
{\Large \bf 
Quasi-localized wavefunctions on magnetized tori \\
and tiny neutrino Yukawa couplings\\*[20pt] } }

\author{
Keigo~Sumita\footnote{
E-mail address: k.sumita@aoni.waseda.jp}\\*[20pt]
{\it \normalsize 
Department of Physics, Waseda University, 
Tokyo 169-8555, Japan} \\*[50pt]}

\date{
\centerline{\small \bf Abstract}
\begin{minipage}{0.9\linewidth}
\medskip 
\medskip 
\small 
This paper shows that, a quasi-localization of wavefunctions in toroidal compactifications with 
magnetic fluxes can lead to a strong suppression for relevant Yukawa couplings, 
and it is applicable to obtain tiny neutrino masses. 
Although it is known that magnetic fluxes lead to a Gaussian profile of 
zero-modes on a torus and that can yield a suppressed coupling 
in higher-dimensional supersymmetric Yang-Mills (SYM) theories, 
the largest (diagonal) entry of Yukawa matrices is always of $\mathcal O(1)$. 
In this paper, we propose a way to induce an absolutely tiny global factor of Yukawa matrices. 
In two SYM theories defined in different dimensional spacetime, 
their bifundamental representations must be localized as a point in some directions. 
Overlaps of such point-like localized wavefunctions and Gaussian zero-modes 
give a global factor of Yukawa matrices, 
and it can be a strong suppression factor or a usual $\mathcal O(1)$ factor, 
corresponding to their distance. 
Our numerical analysis shows that 
it is possible to obtain a suppression strong enough to realize 
the tiny neutrino masses without a hard fine-tuning. 
Furthermore, we propose a concrete model of the tiny neutrino Yukawa couplings 
in a magnetized SYM system. 
\end{minipage}}

\begin{titlepage}
\maketitle
\thispagestyle{empty}
\clearpage
\tableofcontents
\thispagestyle{empty}
\end{titlepage}

\renewcommand{\thefootnote}{\arabic{footnote}}
\setcounter{footnote}{0} 

\section{Introduction} 
Extra dimensional space is an attractive candidate for new physics beyond 
the standard model (SM). This has been being studied very actively 
since Kaluza-Klein theory was proposed to unify the forces of nature. 
Although our circumstances has got more complicated than those at that time, 
such a unified theory is expected to contain the extra dimensional space 
which is an origin of many flavors and the forces of the SM. 
Indeed, superstring theories which are candidates for the unified theory 
is defined in ten-dimensional spacetime, and 
the existence of extra dimensional space is indispensable for 
realizing the flavor structure in those theories. 
Higher-dimensional supersymmetric Yang-Mills (SYM) theories are expected as 
well-motivated higher-dimensional field theories 
since they appear in low-energy limits of superstring theories. 
They are also attractive from the phenomenological viewpoint 
even if we do not consider superstring theories. 
The reason is that they are the most simple theories containing 
supersymmetry (SUSY) which is another great candidate 
for the new physics as well as the extra dimensional space, 
and the SM can be embedded into the SYM theories with a large gauge group.

Toroidal compactifications with magnetic fluxes are able to yield 
the three forces and the flavor structure of the SM in superstring theories and 
higher-dimensional SYM theories\cite{Bachas:1995ik,Cremades:2004wa}. 
Magnetic fluxes on two-dimensional (2D) tori lead to generations of 
the quarks and the leptons, and the Higgs fields as degenerate zero-modes. 
Zero-mode equations for these fields can be solved analytically 
thanks to simplicity of tori, and zero-mode wavefunctions are obtained 
as solutions of the equations. 
We see from the wavefunctions that 
the zero-modes are quasi-localized on the magnetized torus with a Gaussian profile. 
When zero-modes are localized far away from each other, 
their four-dimensional (4D) effective coupling is suppressed\cite{ArkaniHamed:1999dc}. 
Thus, magnetic fluxes on tori can induce hierarchical Yukawa couplings, 
and the hierarchical masses and mixing angles of the quarks and the leptons 
could be potentially obtained on the magnetized torus. 
Indeed, a semi-realistic pattern of them was realized in Ref.~\cite{Abe:2012fj}.

The tiny neutrino mass is a remaining mystery of the SM. 
Although the magnetized toroidal compactifications can yield 
suppressed Yukawa couplings, 
the largest diagonal entry of Yukawa matrices is always estimated to be of $\mathcal O(1)$, 
and another mechanism to realize the tiny masses is needed. 
An elegant solution for explaining the tiny masses is given by 
introducing heavy Majorana mass terms for the right-handed neutrinos, 
which is a so-called seesaw mechanism\cite{seesaw}. 
However, there is a problem in toroidal magnetized compactification of SYM theories. 
Such a Majorana mass term cannot be obtained straightforwardly in the theories. 
We are forced to assume that such mass terms might be generated 
by nonperturbative effects\cite{Ibanez:2006da,Cvetic:2007ku} 
and/or higher-dimensional operators. 
Although there is a wide variety of the seesaw mechanisms, 
they all require the presence of additional mass terms. 
Thus, this paper aim to propose a simple way to realize tiny Yukawa couplings 
for the neutrinos in the magnetized toroidal compactifications.

This paper is organized as follow. 
Section 2 is devoted to review the magnetized toroidal compactifications in 
SYM theories. Zero-mode wavefunctions on the magnetized torus and 
their leading Yukawa couplings are shown. 
We give a main idea and result of this paper in Section 3. 
A numerical analysis is also performed to verify consistency with 
the tiny neutrino masses. 
In Section 4, we propose a concrete model with a specific configuration 
of magnetic fluxes on the tori in a SYM system, 
where our mechanism to yield the tiny neutrino Yukawa couplings will work correctly. 
We conclude with a discussion of prospects in Section 5.

\section{Toroidal compactification with magnetic fluxes} 
This section gives an overview of magnetized toroidal compactifications 
on the basis of six-dimensional (6D) SYM theories compactified on a two-dimensional (2D) torus, 
following Ref.~\cite{Ibanez:2006da} (See also \cite{Ibanez:2012zz} for a review of 
magnetized toroidal compactifications). 
In this section, we use complex coordinates $z$ 
and a complex 2D vector field $A_z$ defined as 
\begin{eqnarray*}
z&\equiv&\frac12\left( x^1+\tau x^2 \right) , \qquad \bar z \equiv \left( z\right)^*,\\
A_z&\equiv&-\frac1{\im\tau}\left(\tau* A_4-A_5\right),\qquad 
\bar A_{\bar z} \equiv \left(A_z\right)^\dagger, 
\end{eqnarray*}
where $(x^1,x^2)$ are the real torus coordinates and 
$(A_4,A_5)$ are extra dimensional components of a (real) 6D vector field. 
Periodicity of this torus is expressed by identifications of $z\sim z+1$ and $z\sim z+\tau$. 
Complex structure of this torus is denoted by $\tau$. 
This is defined with the torus radius $R$ in the torus metric as 
\begin{equation*}
ds^2 = g_{ab}dx^a dx^b  \equiv 2h~ dz d\bar z, 
\end{equation*}
where 
\begin{equation*}
g =\left(2\pi R\right)^2
\begin{pmatrix}
1 & \re\tau\\
\re\tau& |\tau|^2
\end{pmatrix},\qquad 
h=2\left(2\pi R\right)^2. 
\end{equation*}

On this torus, we consider the following gauge potential containing magnetic fluxes and 
Wilson lines, 
\begin{equation*}
A_z = \frac{\pi}{\im\tau}\left( M\bar z +\zeta \right). 
\end{equation*}
The magnetic fluxes and Wilson lines are represented by ($N\times N$)-matrices 
in $U(N)$ SYM theories. 
Since this paper concentrates on Abelian forms of magnetic fluxes and Wilson lines, 
the matrices take nonvanishing values only for their diagonal entries. 
Note that the nonvanishing entries of magnetic fluxes are forced to be integer because of 
the Dirac's quantization condition. 

We discuss gauge symmetry breaking due to these magnetic fluxes and Wilson lines. 
In the case of $U(2)$ SYM theory, 
this $U(2)$ gauge symmetry is broken down to $U(1)\times U(1)$ 
when the two diagonal elements of magnetic fluxes (Wilson lines) are 
chosen to be different values. 
This is extended to the case of $U(N)$ SYM theories, 
and the gauge symmetry is broken as $U(N)\rightarrow\prod_a U(N_a)$, 
where $U(N_a)$ is an unbroken gauge subgroup of $U(N)$. 

We consider zero-mode equations for a two-component spinor field 
on this magnetized torus. Fields living in extra dimensional space are 
decomposed into the zero-mode and multiple Kaluza-Klein modes, 
and we focus on the zero-mode in the following. 
The two-component spinor is denoted by 
\begin{equation*}
\psi=
\begin{pmatrix}
\psi_+\\
\psi_-
\end{pmatrix}, 
\end{equation*}
and then, their zero-mode equations are given by 
\begin{eqnarray*}
\bar\partial_z\psi_+ + [\bar A_{\bar z}, \psi_+ ] &=&0,\\
\partial_z\psi_-  -[A_z, \psi_- ] &=&0.\\
\end{eqnarray*}

In the assumption of gauge symmetry breaking $U(N)\rightarrow\prod_a U(N_a)$, 
a bifundamental representation $(N_a, \bar N_b)$ of two remaining gauge subgroups 
$U(N_a)\times U(N_b)$ contained in $\psi_\pm$  is denoted by $\psi_\pm^{ab}$. 
Their zero-mode equations are given by 
\begin{eqnarray}
\left[\bar\partial_z+ 
\frac{\pi}{2\im\tau}\left( M_{ab}  z +\zeta_{ab} \right)\right] \psi_+^{ab} &=&0,
\label{eq:zeroii}\\
\left[\partial_z - 
\frac{\pi}{2\im\tau}\left( M_{ab}  \bar z +\bar\zeta_{ab} \right)\right] \psi_-^{ab} &=&0,
\label{eq:zeroij}
\end{eqnarray}
where $M_{ab}\equiv M_a-M_b$ and $\zeta_{ab}\equiv \zeta_a - \zeta_b$, 
and $M_a$ and $\zeta_a$ are relevant entries of $M$ and $\zeta$.

These equations were studied in detail in Ref.~\cite{Cremades:2004wa}. 
Either the zero-mode equations for $\psi^{ab}_+$ or $\psi^{ab}_-$ 
gives a well-defined wavefunction, 
which is related to the sign of magnetic fluxes $M_{ab}$. 
In the case of $M_{ab}>0$, 
a bifundamental $\psi^{ab}_+$ has normalizable solutions 
in the zero-mode equation (\ref{eq:zeroii}), 
and then, zero-modes of $\psi^{ab}_-$ are eliminated. 
There appear only the zero-modes of $\psi^{ab}_+$ 
with a degeneracy given by $M_{ab}$. 
In contrast, the zero-mode equation (\ref{eq:zeroij}) yields 
the zero-modes of $\psi^{ab}_-$ when $M_{ab}<0$, 
and its degeneracy is given by $|M_{ab}|$. 
This is a kind of chirality projection to yield a 4D chiral spectrum like the SM. 
Furthermore, 
we can identify these degenerate zero-modes with the generations of the SM particles. 
$|M_{ab}|=3$ unit of magnetic flux leads to the three generations of 
quarks and leptons.

Their 4D effective coupling constants are of our main interest 
because they would be Yukawa couplings of the SM which determine 
the masses and mixing angles of the quarks and leptons. 
The coupling constant of 4D effective field theories is 
given as an overlap integral of wavefunctions on the torus. 
Thus, we are required to study the zero-mode wavefunctions 
obtained as solutions of Eqs.~(\ref{eq:zeroii}) and (\ref{eq:zeroij}). 
The zero-mode wavefunctions are obtained analytically, which are 
expressed by using Jacobi-theta functions. 
We use an index $I_{ab}$ to label the $|M_{ab}|$ degenerate zero-modes, 
that is, $I_{ab}=1,2,\ldots, |M_{ab}|$. 
The wavefunction of the $I_{ab}$-th zero-mode $\psi_+^{I_{ab}}$ for $M_{ab}>0$ 
is then given by 
\begin{eqnarray}
\psi_+^{I_{ab}} &=& \Theta^{I_{ab},M_{ab}}\left(z+\zeta_{ab}/M_{ab}, \tau \right),\nonumber\\
\Theta^{I,M}\left(z,\tau \right) &=& \mathcal N_M  e^{\pi i M z\im z / \im\tau}
\vartheta\begin{bmatrix}
I/M \\ 0 \end{bmatrix}
\left(Mz, M\tau\right), \label{eq:wavefunc}
\end{eqnarray}
where the Jacobi-theta function is defined by 
\begin{equation}
\vartheta
\begin{bmatrix}
a\\b\end{bmatrix}
\left(\nu, \tau\right) = \sum_{ \ell\in \mathbb Z} 
e^{\pi i \left(a+ \ell \right)^2\tau}
e^{2\pi i \left(a+ \ell \right)\left(\nu +b\right)}.
\label{eq:jacobi}
\end{equation}
Normalization factors are determined as 
\begin{equation*}
\int  dzd\bar z \sqrt{{\rm det}\, g} \,\Theta^{I,M} \left(\Theta^{J,M}\right)^* =\delta_{IJ}, 
\end{equation*}
and we find 
\begin{equation}
\mathcal N_M = \left(\frac{2\im\tau|M|}{\mathcal A^2}\right)^{1/4}, \label{eq:normal}
\end{equation}
where $\mathcal A$ denotes the area of torus. 
For $M_{ab}<0$, 
the wavefunction of $\psi_-^{I_{ab}}$ is given by 
\begin{equation*}
\psi_-^{I_{ab}} = \left( 
\Theta^{I_{ab},M_{ab}}\left(z+\zeta_{ab} / M_{ab}, \tau \right)
\right)^*. 
\end{equation*}

\begin{figure}[t]
\begin{center}
\hfil
\includegraphics[width=0.5\linewidth]{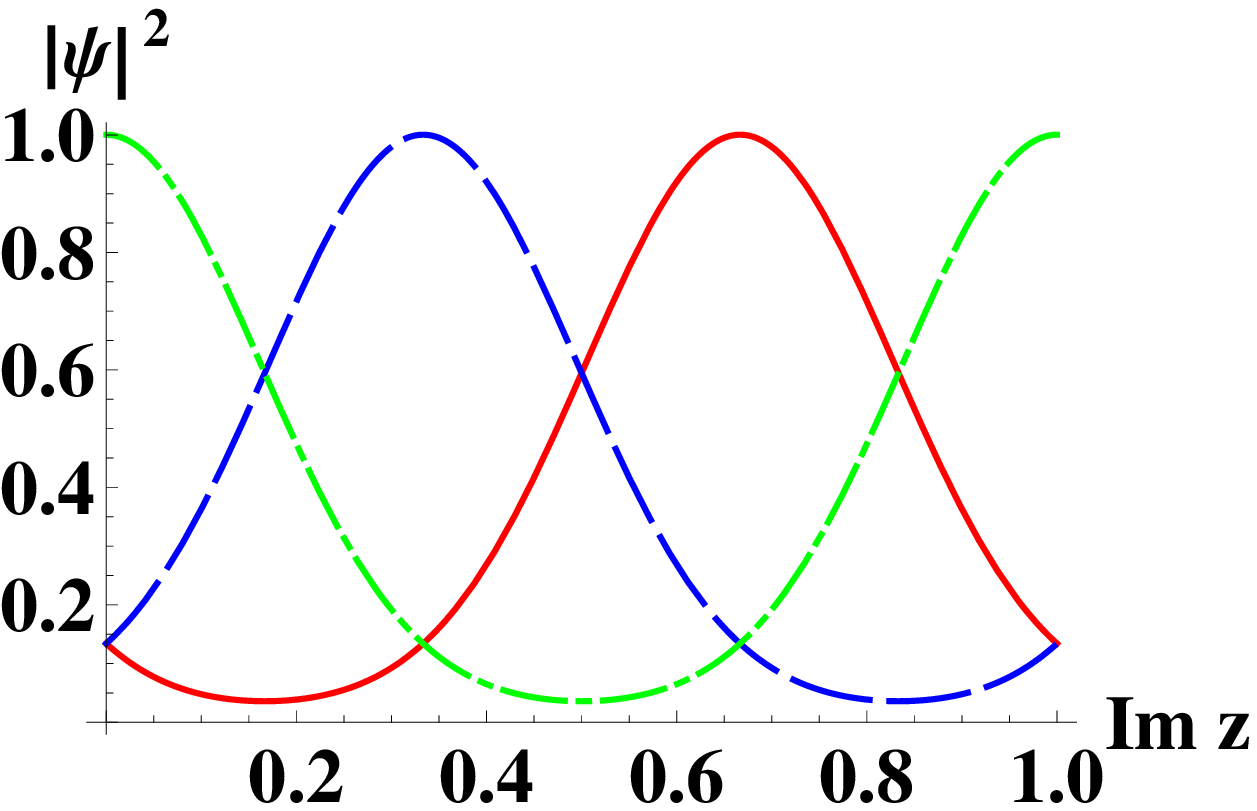}
\hspace{5pt}
\includegraphics[width=0.35\linewidth]{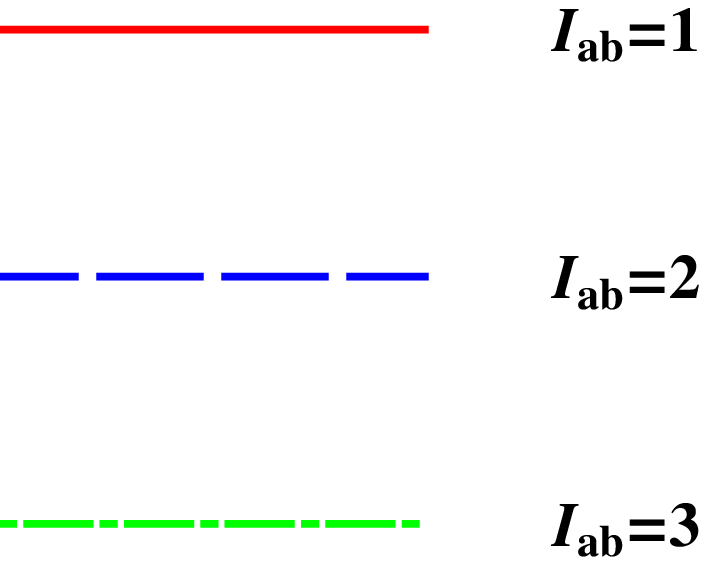}
\caption{The squared absolute values of wavefunctions of the three zero-modes 
$|\psi_+^{I_{ab}}|^2$ given in Eq.~(\ref{eq:wavefunc}) 
with $(M_{ab}=3, \zeta_{ab}=0,\tau = i)$ are shown, 
where we set $\re z=0$ and the horizontal axis corresponds to the other direction of torus. }
\label{fig:threegen}
\end{center}
\end{figure}

These zero-modes have a Gaussian profile in the extra dimensional space. 
For instance, we depict the squared absolute values 
of three wavefunctions $|\psi_+^{I_{ab}}|^2$ given by a parameter set of 
$(M_{ab}=3, \zeta_{ab}=0,\tau = i)$ in a direction of the torus ($\re z=0$) 
in Fig.~\ref{fig:threegen}. 
Three colored lines correspond to the three zero-modes labeled by $I_{ab}=1,2,3$. 
We see from this figure that the degenerate zero-modes are quasi-localized 
at points of  the torus, and their peak positions are different from each other. 
These peak positions can be shifted by introducing 
non-vanishing Wilson lines $\zeta_{ab}\neq0$. 
As we said above, 
Yukawa couplings of the SM are expected to be given by 
an overlap integral of the zero-mode wavefunctions of a left-handed matter field, 
a right-handed matter field and a Higgs field. 
When their wavefunctions are localized far away from each other 
on the magnetized torus, a suppressed Yukawa coupling will be obtained. 
This is able to lead to hierarchical Yukawa couplings. 
Indeed, a semi-realistic pattern of the masses and mixing angles 
of the quarks and the leptons was realized 
without hierarchical input parameters\cite{Abe:2012fj}.

Yukawa couplings can be calculated analytically by performing overlap integrals. 
They originate from the higher-dimensional gauge coupling, 
and the overlap integral is typically given by 
\begin{equation*}
Y_{I_{ab}J_{bc}K_{ca}} = g \int _{T^2} dzd\bar z \,
\Theta^{I_{ab},M_{ab}}\left(z+\zeta_{ab} / M_{ab}, \tau \right)
\Theta^{J_{bc},M_{bc}}\left(z+\zeta_{bc} / M_{bc}, \tau \right)
\Theta^{K_{ca},M_{ca}}\left(z+\zeta_{ca} / M_{ca}, \tau \right), 
\end{equation*}
where $g$ is the higher-dimensional gauge coupling constant. 
Ref.~\cite{Cremades:2004wa} performed this integral and found the result as 
\begin{eqnarray}
Y_{I_{ab}J_{bc}K_{ca}} &=&\frac{\mathcal N_{M_{ab}} 
\mathcal N_{M_{bc}}}{\mathcal N_{M_{ca}}}
\exp\left\{{\frac{\pi i}{\im\tau}\left[\zeta_{ab}\im\frac{\zeta_{ab}}{M_{ab}}
+\zeta_{bc}\im\frac{\zeta_{bc}}{M_{bc}}
+\zeta_{ca}\im\frac{\zeta_{ca}}{M_{ca}}\right]}\right\}\nonumber\\
&&\times\sum_{\ell\in\mathbb{Z}_{M_{ac}}}
\vartheta \begin{bmatrix}
\frac{M_{bc}I_{ab}-M_{ab}J_{bc}+M_{ab}M_{bc}\ell}{M_{ab}M_{bc}M_{ac}}\\ 
0\end{bmatrix} (M_{bc}\zeta_{ab}-M_{ab}\zeta_{bc} , \tau M_{ab}M_{bc}M_{ac})
\nonumber\\ 
&&\qquad\qquad\times\delta_{I_{ab}+J_{bc}+M_{ab}\ell, K_{ca}} , 
\label{eq:yukawa}
\end{eqnarray}
for $M_{ab},M_{bc}>0$ and $M_{ca}<0$. 
In evaluating this result numerically, 
we see that the Yukawa couplings can be hierarchical enough to 
realize the ratio of masses of the top and up quarks ($\sim\mathcal O (10^{-5})$). 

The largest entry of $Y_{I_{ab}J_{bc}K_{ca}}$, 
which is identified with a $(3,3)$ entry of Yukawa matrices, 
is obtained to be of $\mathcal O(1)$ invariably. 
One can easily see that, because the dominant contribution is given 
by $a=0$ and $\ell=0$ in the Jacobi-theta function (\ref{eq:jacobi}). 
Thus, tiny neutrino mass ($\nu \lesssim \mathcal O(1{\rm eV})$) 
cannot be obtained with these Yukawa couplings, 
without heavy Majorana masses for the right-handed neutrinos\cite{seesaw}. 
(The top/bottom ratio can be realized by using the degree of freedom of 
$\tan\beta\equiv \langle H_u\rangle / \langle H_d\rangle$, where 
$H_u$ ($H_d$). represents the up- (down-) type Higgs field)
This paper provides a way to yield the absolutely tiny Yukawa couplings 
in the magnetized toroidal compactifications, 
which has a global strong suppression factor.

\section{Suppression factor in mixed SYM systems}
This section explains the basic idea for realizing 
a global factor to strongly suppress the Yukawa couplings. 
We consider a mixture of two SYM theories defined in different dimensions of spacetime, 
especially, we focus on a system consisting of 6D and 10D SYM theories which is 
well motivated by stable D-brane systems in type IIB superstring theory. 
These mixture configurations generically contain bifundamental representations 
charged under both the 6D and 10D SYM theories. 
Zero-mode wavefunctions of such bifundamentals depends on 
all the 10D coordinates, 
but they must be localized as a point in four extra directions where 
the 6D SYM fields cannot move. 

We consider a Yukawa coupling among the two bifundamental fields and 
an adjoint field of 10D SYM theory. The two bifundamental fields should be 
localized at a point in the four direction but the 10D adjoint field has a Gaussian profile 
on the magnetized tori. 
This Yukawa coupling in the 4D effective theory 
is determined by overlap integrals of zero-mode wavefunctions on each of the three 2D tori. 
The first one is given on the torus where the 6D fields live, 
and this part is represented by Eq.~(\ref{eq:yukawa}). 
The magnetic fluxes on this torus will induce the hierarchical generation structure. 
On the other two tori, the zero-mode wavefunctions of 10D field is still expressed by 
Eq.~(\ref{eq:wavefunc}), 
but the zero-modes of the other two bifundamental representations are not. 
The point-like localization of the bifundamental fields should be expressed 
by a delta function.  
Introducing the delta function to express the localizations as a point 
seems to be also a sensible proposal 
in a D-brane picture of superstring theories\cite{Cremades:2004wa}. 
Thus, the integration on the torus is typically given by \cite{Cremades:2004wa,Abe:2015jqa}
\begin{equation}
g\int dz d\bar z\, \delta_{T^2} (z +\chi) \Theta^{I_{ab},M_{ab}}\left(z+\zeta_{ab}/M_{ab}, \tau \right) 
= g\,\Theta^{I_{ab},M_{ab}}\left(-\chi+\zeta_{ab}/M_{ab}, \tau \right), 
\label{eq:supfactor}
\end{equation}
where the delta function $\delta_{T^2} (z +\chi)$ must satisfy the periodicity of torus, 
and $\chi$ indicates the localization point. 
The index $I_{ab}$ labels $|M_{ab}|$ flavors induced by magnetic fluxes on this torus, 
and this factor is universal for generations originating from the other tori. 

\begin{figure}[t]
\begin{center}
\includegraphics[width=0.4\linewidth]{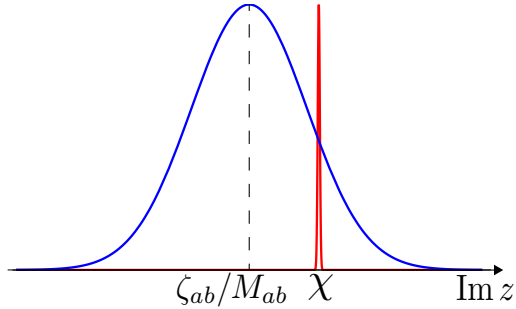}
\caption{A schematic figure of the overlap integral (\ref{eq:supfactor}) is shown. 
The blue Gaussian represents the wavefunction of 10D field and the red line 
expresses the point-like localization of 6D fields. }
\label{fig:image}
\end{center}
\end{figure}

This overlap (\ref{eq:supfactor}) is schematically depicted in Fig.~\ref{fig:image}. 
This integration can take a tiny value when a peak position of 
$\Theta^{I_{ab},M_{ab}}\left(z+\zeta_{ab}/M_{ab}, \tau \right)$ is far away from 
a point of magnetized torus indicated by $\chi$. 
That is, the distance $|\chi-\zeta_{ab}/M_{ab}|$ determines the magnitude of overlap. 
Furthermore, this suppression factor depends on the distance exponentially, and thus, 
we can obtain very strong suppression even when 
the value of distance $|\chi-\zeta_{ab}/M_{ab}|$ is of $\mathcal O(1)$.

We evaluate the suppression factor to verify consistency with neutrino masses, 
which roughly requires an $\mathcal O(10^{-10})$ suppression\cite{Agashe:2014kda}. 
In generic ($4+2n$)-dimensional SYM theory compactified on magnetized tori, 
the net number of degenerate zero-modes of a bifundamental representation 
is given by a product of relevant magnetic fluxes on each torus, 
\begin{equation*}
N_{ab} =\prod_{i=1}^n  \left|M_{ab}^{i}\right|. 
\end{equation*}
Three-generation structure of matter fields must originate from a single torus, 
and then, 
magnetic fluxes on the other tori are required not to generate extra zero-modes 
(i.e., $N_{ab} = 3\times1\times1$). 
Furthermore, all the generation structure of the SM must originate from the same single 
torus, otherwise the rank of Yukawa matrices is to be reduced to one and 
some of matter fields would remain massless after the electroweak symmetry breaking.

We consider a mixture of 
a 6D theory compactified on a magnetized torus $(T^2)_1$ 
and a 10D SYM theory on magnetized $(T^2)_1\times (T^2)_2\times (T^2)_3$ in the following. 
When all the three generation structure of the SM is obtained on the first torus $(T^2_1)$, 
the other magnetized torus should not induce generations. 
This set-up implies $M_{ab}=1$ and $I_{ab}=1$ in the suppression factor (\ref{eq:supfactor}) 
given on the second and the third tori. 
Such configurations can be indeed exist, and 
we will give a concrete model with such a configuration preserving 
$\mathcal N=1$ SUSY in the next section.

The suppression factor (\ref{eq:supfactor}) on the $i$-th torus 
is a function of two parameters, $\tau^i$ and $X^i\equiv\zeta^i_{ab}/M^i_{ab}-\chi^i$. 
Note that, $\tau^i$ denotes the complex structure of the $i$-th torus, and 
$X^i$ represents a distance between the peak position of Gaussian-like wavefunction 
of a 10D field and 
a point at which other bifundamental fields are strongly localized by the delta function. 
In the following analysis, we neglect the gauge coupling constant $g$ and 
the normalization factor (\ref{eq:normal}) contained in the function $\Theta^{I,M}$, 
because their product 
\begin{equation*}
g\mathcal N_1 =\left(2\im\tau\right)^{1/4}e^{\phi/2}\sqrt{\frac{\alpha'}{\mathcal A}},
\end{equation*}
is always estimated as an $\mathcal O(1) $ factor\footnote{
We may choose a slightly large value of $\im\tau \sim \mathcal O(10)$ 
to get a sufficiently strong suppression, 
but it would not affect on this discussion because 
this neglected prefactor contains only the fourth root of it. } 
as long as the compactification scale is as high as 
$M_{\rm GUT}=2.0 \times10^{16}$. 
Through these, the suppression factor (\ref{eq:supfactor}) is rewritten as
\begin{equation}
f^i(X^i,\tau^i)\equiv e^{\pi i X^i\im X^i / \im\tau^i}\sum_{ \ell\in \mathbb Z} 
e^{\pi i \ell^2\tau^i}
e^{2\pi i \ell X^i },\label{eq:supff}
\end{equation}
where we use the following property of wavefunction, $\Theta^{1,1} = \Theta^{0,1}$.

Our numerical analysis of the function (\ref{eq:supff}) reveals that 
this factor can lead to tiny neutrino masses without strong fine tuning, 
which is shown in Table \ref{tb:suppress}. 
\begin{table}[th]
\center
\begin{tabular}{c|cc}
 & case1 & case2\\ \hline\hline
$\im X^2$&$15$&$7.5$\\
$\im\tau^2$&$30$&$15$\\
$f^2(X^2,\tau^2)$& $1.2\times10^{-10}$ &$1.5\times 10^{-5}$\\\hline
$\im X^3$&0&$7.5$\\
$\im \tau^3$&$1.0$&$15$\\
$f^3(X^3,\tau^3)$&$1.1$ & $1.5\times 10^{-5}$\\\hline
$f^2\times f^3$&$1.3\times10^{-10}$&$2.3\times10^{-10}$
\end{tabular}
\caption{ Function(\ref{eq:supff}) is evaluated in two typical cases, which 
shows that the sufficient suppression to yield tiny neutrino masses is obtained with 
$\mathcal O(10)$ values of input parameters. }
\label{tb:suppress}
\end{table}
The suppression can be caused on the second and the third tori, 
and the 4D effective couplings contain their product, $f_2\times f_3$. 
In the case1, the second torus generates the tiny value of $f_2$ and the other 
is of $\mathcal O(1)$. 
On the other side, both functions, $f_2$ and $f_3$, contribute evenly, 
where the value of input parameters become mild. 
In both cases, the values of input parameters are not so sensitive. 
With deviations of $(\pm0.5)$ from the sample values shown in the table, 
almost the same result would be obtained. 
The real parts of $X^i$ and $\tau^i$ just induce unphysical phases in this factor.

\section{A concrete model}
We construct a specific model of the tiny neutrino Yukawa couplings 
respecting the minimal supersymmetric standard model (MSSM), 
on the basis of 6D and 10D SYM theories compactified on magnetized tori. 
A mixture of 6D $U(1)$ gauge theory compactified on $(T^2)_1$ and 
10D $U(7)$ SYM theory on $(T^2)_1\times (T^2)_2\times (T^2)_3$ is our starting point. 
We parametrize their magnetic fluxes as 
\begin{eqnarray*}
M_{6D} &=& m,\\
M^{i}_{10D} &=& \begin{pmatrix} 
M^i_C\times {\bm1}_3&0&0\\
0&M^i_L\times {\bm1}_2&0\\
0&0&M^i_R\times {\bm1}_2
\end{pmatrix},
\end{eqnarray*}
where $M_{6D}$ represents the magnetic flux on the first torus in the $U(1)$ gauge theory, 
and $M^i_{10D}$ expresses those on the $i$-th torus in the $U(7)$ SYM theory. 
In Ref.~\cite{Cremades:2004wa,Troost:1999xn}, 
conditions to preserve $\mathcal N=1$ SUSY out of the full $N=2,3$ and 4 in terms of 
the 4D supercharges were studied, and it is given by
\begin{equation}
m=0,\qquad \sum_{i=1}^3\frac{M^{i}_{10D}}{\mathcal A^{i}} = 0, \label{eq:susucon}
\end{equation}
where $A^{i}$ is the area of the $i$-th torus. 
This form of the magnetic fluxes generically break the gauge symmetry as 
$U(7)\rightarrow U(3)_C\times U(2)_L\times U(2)_R$. 
The last $U(2)_R$ gauge should be further broken to $U(1)_{R'}\times U(1)_{R''}$ 
by introducing Wilson lines of the form 
\begin{eqnarray*}
\zeta^{i}_{10D} &=& \begin{pmatrix}
\zeta^i_C\times {\bm1}_3&0&0&0\\
0&\zeta^i_L\times {\bm1}_2&0&0\\
0&0&\zeta^i_{R'}&0\\
0&0&0&\zeta^i_{R'}
\end{pmatrix}.
\end{eqnarray*}

In this parametrization, $U(7)$ adjoint representations are decomposed, 
which yield the MSSM fields except for the leptons. 
Bifundamental representations of $U(7)\times U(1)$ accommodates 
the charged lepton multiplets and the neutrino multiplets. 
Note that, the suppression factor is given to Yukawa couplings of the charged leptons 
as well as the neutrinos, but the distance $X^i$ for the each sector 
can be chosen independently. 
Thus, we can set them to give rise to just an $\mathcal O(1)$ factor for the charged leptons 
while realizing tiny neutrino Yukawa couplings. 

We can find a configuration of magnetic fluxes, which 
leads to three generation structure for the quarks and leptons satisfying 
conditions (\ref{eq:susucon}) to preserve $\mathcal N=1$ SUSY, as 
\begin{eqnarray*}
M_{6D} &=& 0,\\
\left(M^1_C,\,M^1_L,\,M^1_R\right) &=& \left(0,~~3,-3\right),\\
\left(M^2_C,\,M^2_L,\,M^2_R\right) &=& \left(0,-1,~~0\right),\\
\left(M^3_C,\,M^3_L,\,M^3_R\right) &=& \left(0,~~0,~~1\right). 
\end{eqnarray*}
The SUSY condition is clearly satisfied with 
$\mathcal A^{1}/\mathcal A^{2} = \mathcal A^{1}/\mathcal A^{3} = 3$. 
This configuration leads to three generations of ($3,\bar 2,1$) and ($\bar 3,1,2$) 
representation of $U(3)_C\times U(2)_L\times U(2)_R$, which can be identified with 
the quark multiplets. Six generations of ($1,2,\bar 2$) representation 
correspond to the Higgs multiplets. 
The existence of multiple Higgs fields is a generic feature of 
magnetized SYM models, and we identify a linear combination of them with 
the MSSM Higgs multiplets. The other five combinations should be decoupled somehow. 
The three generations of the charged lepton and neutrino multiplets 
are correctly given by the bifundamental representation 
charged under the original $U(1)$ gauge group as we mentioned above. 
Thus, Yukawa couplings including the leptons have 
the global suppression factor (\ref{eq:supfactor}). 

On this magnetized background, 
we can obtain 4D effective action in accordance with an elegant way\cite{Abe:2015jqa}, 
and there, specific spectra of the quarks, the leptons, and the SUSY particles 
can also be calculated. 
Although we will not do that because that is not our purpose in this paper, 
we are able to easily infer that similar spectra to those obtained in Ref.~\cite{Abe:2012fj} 
can be realized except for the neutrino sector. 

\section{Conclusions and Discussions} 
We have studied strong suppression due to quasi-localization of wavefunctions 
and realized tiny neutrino Yukawa couplings in magnetized toroidal compactifications 
of SYM theories. 
In a system of two SYM theories defined in different dimensions, such as, 
6D-10D SYM theories, 
bifundamental representations of the two gauge groups must be 
localized as a point in extra four directions, 
which should be represented by the well-defined delta function. 
As a result, 
their 4D effective coupling with a 10D field contains a factor induced by 
overlap integrals of the delta function and a Gaussian wavefunction 
of the 10D field in the four directions. 
When a point on which the delta function will not vanish is equal to 
the peak position of Gaussian, the integral induces an $\mathcal O(1)$ factor. 
However, for their nonvanishing distance, 
the value of the factor is reduced exponentially. 
Our numerical analysis shows that 
the $\mathcal O(10)$ of distance leads to an 
$\mathcal O(10^{-10})$ of suppression, 
which is strong enough to yield tiny neutrino masses consistent with 
experimental and observational data.

A specific SYM system and its configuration have been discussed in Section 4. 
We consider a mixture of a 6D $U(1)$ gauge theory compactified on magnetized $(T^2)_1$ 
and a 10D $U(7)$ SYM theory on $(T^2)_1\times (T^2)_2\times (T^2)_3$. 
The $U(7)$ gauge group is broken in part by magnetic fluxes to 
$U(3)_C\times U(2)_L \times U(2)_R$. 
The decomposed $U(7)$ adjoint field accommodates 
the MSSM contents except for the lepton multiplets. 
The lepton multiplets assigned into the 
bifundamental representation charged under the original (6D) $U(1)$ 
gauge group. 
Thus, Yukawa couplings for the charged leptons and 
the neutrinos are determined by an overlap integral of 
three Gaussian wavefunctions on the first torus $(T^2)_1$, 
and by overlap integrals of the delta function and the Gaussian on 
$(T^2)_2$ and $(T^2)_3$. 
Since the distance $X^i$ between the peak of the delta function and the Gaussian 
can be controlled independently for the charged leptons and the neutrinos, 
neutrino Yukawa couplings can be strongly suppressed while 
just an $\mathcal O(1)$ factor is given rise to for the charged lepton sector.

This very strong suppression could be applied to other couplings 
to forbid dangerous processes, such as, the proton decay and the flavor violation. 
In magnetized toroidal compactifications, higher-order couplings can also be 
calculated analytically\cite{Abe:2009dr}, and 
this strong suppression mechanism would also work 
in such higher-order couplings 
as well as Yukawa couplings. 
Thus, we expect that a wide variety of its applications are possible. 
We will construct models focusing on other phenomenologies 
and analyze them elsewhere. 



\end{document}